\documentclass[journal]{IEEEtran}

\usepackage{amsmath}
\usepackage{amssymb}
\usepackage{cite}
\usepackage{color}
\usepackage{graphicx}
\usepackage{subfigure}
\usepackage{url}
\usepackage{soul}

\begin{document}
\title{Vehicular Communications: A Physical Layer Perspective}
\author{
Le~Liang,~\IEEEmembership{Student Member,~IEEE},
Haixia Peng,~\IEEEmembership{Student Member,~IEEE},
\\
Geoffrey Ye Li,~\IEEEmembership{Fellow,~IEEE},
and Xuemin (Sherman) Shen,~\IEEEmembership{Fellow,~IEEE}


\thanks{The work of L. Liang and G. Y. Li was supported in part by a research gift from Intel Corporation and the National Science Foundation under Grants 1405116 and 1443894.
The work of H. Peng was supported in part by the National Natural Science Foundation of China under Grant 91638204.
}
\thanks{L. Liang and G. Y. Li are with the School of Electrical and Computer Engineering, Georgia Institute of Technology, Atlanta, GA 30332 USA (e-mail: lliang@gatech.edu; liye@ece.gatech.edu).}
\thanks{H. Peng and X. Shen are with the Department of Electrical and Computer Engineering, University of Waterloo, Waterloo, ON N2L 3G1 Canada (e-mail: h27peng@uwaterloo.ca; xshen@bbcr.uwaterloo.ca).}
}

\maketitle

\begin{abstract}
Vehicular communications have attracted more and more attention recently from both industry and academia due to their strong potential to enhance road safety, improve traffic efficiency, and provide rich on-board information and entertainment services.
In this paper, we discuss fundamental physical layer issues that enable efficient vehicular communications and present a comprehensive overview of the state-of-the-art research.
We first introduce vehicular channel characteristics and modeling, which are the key underlying features differentiating vehicular communications from other types of wireless systems.
We then present schemes to estimate the time-varying vehicular channels and various modulation techniques to deal with high-mobility channels.
After reviewing resource allocation for vehicular communications, we discuss the potential to enable vehicular communications over the millimeter wave bands.
Finally, we identify the challenges and opportunities associated with vehicular communications.
\end{abstract}


\section{Introduction}

\IEEEPARstart{T}{he} emerging vehicular communications, or vehicle-to-everything (V2X) communications, are expected to enable a whole new set of services, ranging from road safety improvement to traffic efficiency optimization, from driverless cars to ubiquitous Internet access on vehicles \cite{Karagiannis2011vehicular,Papadimitratos2009vehicular,3GPPr14v2xstudy,Araniti2013lte}.
They hold significant potential in making our daily experience on wheels safer and more convenient.

In recent years, various communication standards have been developed across the globe to ensure interoperability in information exchange of vehicles, e.g., dedicated short-range communications (DSRC) standards in the US \cite{Kenney2011dsrc,ieee2010ieee} and intelligent transportation system (ITS)-G5 standards developed by the European Telecommunications Standards Institute (ETSI) \cite{ITSG5}. Both standards are based on the IEEE 802.11p technology, which establishes the foundation for communications in vehicular ad hoc networks (VANETs).
However, recent studies \cite{Araniti2013lte,Mir2014lte,hafeez2013performance} show that vehicular communications based on IEEE 802.11p face several challenges, such as short-lived vehicle-to-infrastructure (V2I) connections, potentially unbounded channel access delay, and lack of quality-of-service (QoS) guarantee, due to its physical and medium access control (MAC) layer designs that have been originally optimized for wireless local area networks (WLAN) with low mobility.

\begin{figure}[!t]
\centering
\includegraphics[width=0.46\textwidth]{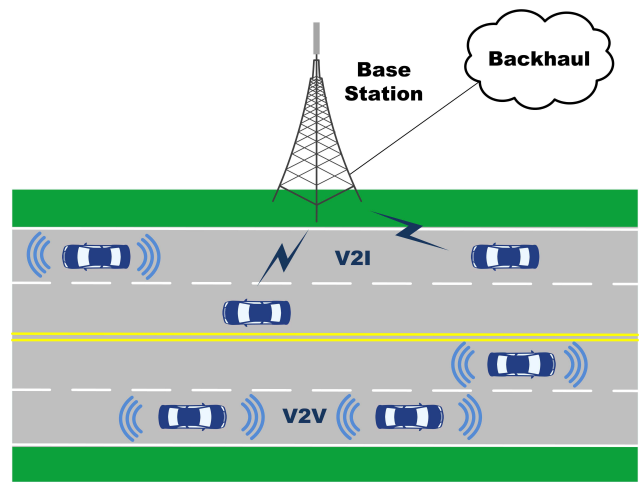}
\caption{An illustration of vehicular communication systems.}\label{fig:v2x}
\end{figure}

Recently, the 3rd Generation Partnership Project (3GPP) has also started looking into supporting V2X services in cellular networks \cite{3GPPr14v2xstudy,Sun2016support,Schwarz2017signal,Seo2016LTE}.
For example, telecommunication and automotive leaders have founded the 5G automotive association (5GAA), a cross-industry consortium, to actively push the development and deployment of cellular-based V2X communications.
Widely deployed cellular networks, assisted with direct device-to-device (D2D) underlay communications \cite{Araniti2013lte,Feng2013device,Qiao2015enabling}, potentially provide a promising solution to enable efficient and reliable vehicle-to-vehicle (V2V) and V2I communications, to meet demanding V2X requirements and provide immunity to high mobility due to several intrinsic advantages.
First, cellular networks exercise flexible centralized control over network resources, such as fast link adaptation and dynamic user scheduling, which guarantee optimal network performance.
Second, the large capacity and proved maturity of cellular networks can provide reliable support for a wide variety of bandwidth-thirsty applications and also ease the implementation of V2X communications.
Finally, the side D2D links, complementing the centralized cellular architecture, will provide direct local message dissemination with substantially reduced latency, thus suitable for delay-sensitive V2V communications. Meanwhile, existence of the always-on base station can potentially be beneficial to communications among vehicles through providing side information to the V2V links.

Recognizing the pros and cons of IEEE 802.11p and cellular based vehicular communications, there has been an increasing interest in studying interworking of these two technologies to form a converged network.
For example, in \cite{Abboud2016interworking} and \cite{Zheng2015heteterogeneous}, hybrid DSRC-cellular vehicular network structures have been proposed to meet the requirements of heterogeneous vehicular applications and the associated challenges, such as vertical handover and optimal network selection, have been discussed.

Over the past decade, there have been some excellent review papers on the topic of vehicular communications \cite{Papadimitratos2009vehicular,Karagiannis2011vehicular,Araniti2013lte,Abboud2016interworking,Zheng2015heteterogeneous}.
They have identified potential applications of vehicular networks and the requirements on vehicular network design.
The associated challenges, such as extremely low latency and high reliability provisioning, heterogeneity of data dissemination, geo-casting, and security and privacy issues, have been introduced and the corresponding potential solutions, mainly centered around the networking and MAC designs, have been summarized.

This survey differentiates itself from the existing literature in that it is dedicated to understanding various fundamental issues of vehicular communications from a physical layer perspective, which is instrumental in providing solutions based on the underlying physical mechanism and laying the foundation for upper layer designs.
Specifically, we discuss basic physical layer issues, such as vehicular channel characterization and modeling, channel parameter estimation for time-varying high-mobility environments, and modulation design with special consideration of the rapid temporal variation exhibited by vehicular channels.
We also present an overview on the emerging topics in vehicular communications, such as the design of resource allocation schemes for D2D-enabled vehicular networks with minimal channel state information (CSI) requirements and exploiting the millimeter wave (mmWave) bands for broadband vehicular communications.
This comprehensive review on the state-of-the-art research in various physical layer issues of vehicular communications helps identify critical challenges and research opportunities that warrant further investigation and hold great potential in substantially improving the performance of the vehicular communications systems.

The rest of this paper is organized as follows.
In Section~\ref{sec:modeling}, we introduce typical wireless channel characterization methods and various models for vehicular channels.
In Section~\ref{sec:estimation}, we discuss state-of-the-art vehicular channel estimation approaches accounting for high Doppler spread.
In Section~\ref{sec:modulation}, we review modulation designs to deal with interchannel interference (ICI) in multicarrier transmission due to frequency dispersion in high mobility vehicular environments.
In Section~\ref{sec:resrouce}, we address resource allocation for vehicular communications.
Then in Section~\ref{sec:mmWave}, we explore potential in enabling vehicular communications over the mmWave bands for broader bandwidth and higher data rate.
Finally, challenges and opportunities for vehicular communications are identified in Section~\ref{sec:challenges} and conclusions are drawn in Section~\ref{sec:conclusion}.


\section{Vehicular Channels}\label{sec:modeling}
Special channel propagation characteristics are the most fundamental differentiating factor of vehicular communications compared with other types of wireless systems.
Vehicular channels exhibit rapid temporal variability and inherent non-stationarity of channel statistics due to their unique physical environment dynamics \cite{Viriyasitavat2015vehicular,Bernado2014delay}.
As a result, proper mathematical modeling for vehicular channels is vital in evaluating their impacts on vehicular system design and analyzing network performance.
This section is dedicated to characterizing vehicular channels, which will help design various physical layer techniques to enable reliable and efficient vehicular communications.

\subsection{Wireless Channel Characterization}
To model vehicular channels, key parameters need to be properly defined and understood first.
Wireless signals propagate from the transmitter to the receiver through different paths, each with different time-varying propagation delays, amplitude attenuations, and phase rotations. These multipath components may then add constructively or destructively at the receiver, leading to the so called \emph{fading} effect. Such effect can be roughly divided into two types, i.e., large-scale fading and small-scale fading, depending on their impact distances or periods.

One component of the large-scale fading effect is pathloss, which characterizes the average power reduction of a radio signal due to space loss, absorption, diffraction, etc.
Large-scale channel effect also includes shadowing, caused by obstacles between the transmitter and the receiver attenuating signal power through absorption, reflection, scattering, to name a few.
The combined effects of pathloss and shadowing can be characterized in the following simplified expression \cite{Goldsmith2005wireless}
\begin{align}
P_t - P_r  = K + 10\gamma\log_{10}\frac{d}{d_0} + \phi_{\text{dB}}, \quad (\text{in dB})
\end{align}
where $P_t$ and $P_r$ denote the transmitted and the received power levels in dB, respectively, $K$ is the power attenuation at the reference distance $d_0$, $\gamma$ is called the pathloss exponent, $d$ is the distance between the transmitter and the receiver antennas, and $\phi_{\text{dB}}$ in dB is with Gaussian distribution, which represents the shadowing effect and can be modeled as a log-normal random variable.

Small-scale fading takes place over the distance comparable to the carrier wavelength due to the constructive and destructive effects of multiple time-varying delayed replicas of the original signals received from different paths between the transmitter and the receiver. It is traditionally characterized by the time-varying \emph{channel impulse response} \cite{Steele1992mobile}
\begin{align}\label{eq:impulse}
h(\tau, t) = \sum\limits_{i}\alpha_i(t)\delta(\tau-\tau_i),
\end{align}
where $\tau_i$ is the delay of the $i$th path and $\alpha_i(t)$ is the time-varying path attenuation with $\mathbb{E}\left[|\alpha_i|^2\right]=\sigma_i^2$, which can be modeled by a wide-sense stationary (WSS) random process and is independent for different paths.

Channel power delay profile (PDP) \cite{Goldsmith2005wireless}, specified by the delay, $\tau_i$, and the average power, $\sigma_i^2$, of each path, provides an effective way to describe channel's frequency selective characteristics.
The \emph{delay spread}, $\tau_s$, is a compact description of the delay dispersion of a channel and can be derived from the PDP of a channel by
\begin{align}
\tau_s = \sqrt{\frac{\sum\sigma_i^2(\tau_i-\tau_a)^2}{\sum\sigma_i^2}},
\end{align}
where $\tau_a = \sum\sigma_i^2\tau_i/(\sum\sigma_i^2)$.
The delay spread closely relates to the channel frequency selectivity. In particular, the \emph{channel coherence bandwidth}, $B_c$, over which the channel frequency response can be roughly treated as uniform, is inversely proportional to the delay spread, i.e.,
$
B_c \approx c_0/\tau_s
$
up to a constant $c_0$.

On the other hand, Doppler spectrum describes how the path attenuation, $\alpha_i(t)$, changes with time and is defined as
\begin{align}
P_i(f) = \int_{-\infty}^{+\infty} \rho_i(\tau)\exp(-j2\pi f\tau) \text{d} f,
\end{align}
where
$
\rho_i(\tau) = \mathbb{E}\left[ \alpha_i(t+\tau)^*\alpha_i(t) \right].
$
The time variation of vehicular channels can be simply characterized by the \emph{Doppler spread}, which can be derived from the Doppler power spectrum of a channel.
The \emph{channel coherence time} indicates the duration over which the channel impulse response remains roughly constant.
Similar to the relationship between the delay spread and coherent bandwidth, the Doppler spread and the channel coherence time are inversely proportional to each other.

Wireless channels are often assumed to be WSS, which implies the channel statistics do not vary over time. Additionally, channel response at different delays is often assumed to be uncorrelated, leading to the so called uncorrelated scattering (US).
Combining the two assumptions yield the WSSUS channel model, first introduced in \cite{Bello1963characterization}. However, the WSS property may not hold in vehicular channels due to the rapid movement of transmitters and receivers as well as surrounding scatterers, giving rise to, e.g., appearance and disappearance of some scatterers/reflectors \cite{Matolak2008channel}.

Finally, a finite-state Markov chain (FSMC) is sometimes used to simplify fading channel characterization to ease theoretical performance analysis \cite{sadeghi2008finite}. In this model, fading is approximated by a set of finite channel states and varies according to a set of Markov transition probabilities. For instance, a semi-Markov process has been exploited in \cite{Lei2016stochastic} to characterize train mobility in high-speed railway communications systems and the stochastic delay has been analyzed for train control services based on the stochastic network calculus.


\subsection{V2X Channel Modeling}

Depending on the availability of geographic information and the affordable implementation complexity, there are generally three major approaches to model vehicular channels, i.e., deterministic \cite{Maurer2004new,Wiesbeck2007characteristics,Pilosu2011RADII,Nuckelt2013comparison,Guan2016influence}, geometry-based stochastic
\cite{Patzold2008modeling,Zajic2008space,zajac2009three,Zajic2014impact,Cheng2009adaptive,Cheng2009geometry}, and non-geometric stochastic models \cite{Marum2007six,Sen2008vehicle,Wang2012improving}.
Basic principles and representative examples of the three different modeling approaches are presented and discussed below.
For more detailed analyses, comparison of these approaches, and their validation against measurement data, please refer to \cite{Molisch2009survey,Mecklenbrauker2011vehicular} and the references therein.

\subsubsection{Deterministic Models}

\begin{figure}[!t]
\centering
\includegraphics[width=0.46\textwidth]{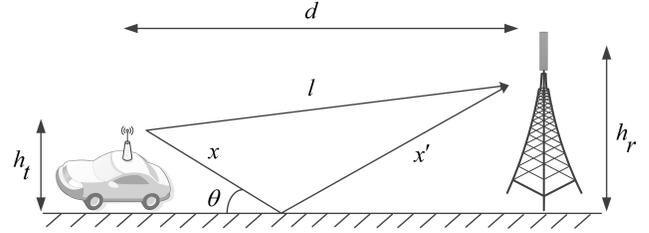}
\caption{A simplified two-ray tracing model for V2I channels \cite{Goldsmith2005wireless}.}\label{fig:ray}
\end{figure}

The deterministic models characterize the channel propagation in a completely deterministic way.
Ray-tracing is a typical example of such a model, which approximates the propagation of electromagnetic waves through solving simplified wave equations given the geometric and dielectric properties of the surrounding scatterers. As a result, ray-tracing provides a site-specific description of the channel parameters for a certain propagation environment and tends to be computationally intensive.

An example of a simplified two-ray V2I channel model is illustrated in Fig.~\mbox{\ref{fig:ray}}, where the received signal consists of a line-of-sight (LoS) ray and a ground-reflected ray. If $s(t)$ is transmitted, the received signal for the model is given by \cite{Goldsmith2005wireless}
\begin{align}
r(t) =& \text{Re}\left\{\frac{\lambda}{4\pi}\left[ \frac{\sqrt{G_l}s(t)e^{-j2\pi l/\lambda}}{l}\right.\right. \nonumber \\
 & \left.\left.  ~~~ + \frac{R\sqrt{G_r}s(t-\tau)e^{-j2\pi(x+x')/\lambda}}{x+x'} \right]e^{j2\pi f_c t} \right\},
\end{align}
where $G_l$ and $G_r$ are the antenna gains of the LoS and reflected rays, respectively, $\tau = (x+x'-l)/c$ is the propagation time difference between the two rays, $R$ is the ground reflection coefficient, $f_c$ and $\lambda=\frac{c}{f_c}$ ($c$ is the speed of light) are the carrier frequency and wavelength, respectively.

Channel models based on three-dimensional ray-tracing for realistic traffic scenarios have been developed in \cite{Maurer2004new,Wiesbeck2007characteristics} and validated with measurements.
A more realistic and scalable method based on ray tracing data interpolation and interfacing has been proposed in \cite{Pilosu2011RADII} to model urban propagation, which provides a way to compute average attenuation to reduce the computational load.
The accuracy of a ray-tracing V2V channel model for urban intersections as compared with channel sounder measurement data has been studied in \cite{Nuckelt2013comparison}, where effects of road-side obstacles, such as traffic signs or parked vehicles, have also been discussed.
Furthermore, the effects of the traffic signs along roads have been incorporated in channel modeling in \cite{Guan2016influence}, where they are decomposed into many small elements to embed the developed analytical models into ray-tracing tools, thus improving propagation modeling accuracy.

\subsubsection{Geometry-based Stochastic Models}
The geometry-based stochastic modeling approach is based on randomly generating the geometry (scatters) of the propagation environment according to certain stochastic distributions together with simplified ray tracing \cite{Molisch2009survey}. The random scatters can be placed in one-ring, two-ring, ellipses, or other shapes, making such a modeling approach fairly flexible. The geometry-based stochastic model also allows the space-time-frequency correlation statistics to be derived \cite{Patzold2008modeling,Zajic2008space}, which can be exploited to augment receiver synchronization and channel estimation.
In \cite{Zajic2008space}, a two-ring geometry-based stochastic model with both single- and double-bounced rays has been proposed for narrowband multiple-input multiple-output (MIMO) V2V Ricean fading channels, where space-time correlation and space-Doppler power spectrum have been derived in a two-dimensional non-isotropic scattering environment.
It has been further extended to a three-dimensional wideband geometric model in \cite{zajac2009three} and to include both stationary and moving scatters around the transmitter and receiver in \cite{Zajic2014impact}.
In \cite{Cheng2009adaptive}, a generic and adaptive geometry-based narrowband model has been proposed for non-isotropic MIMO V2V Ricean fading channels employing combined two-ring and ellipse models, where the impact of the vehicular traffic density has been taken into account.
This has then been extended to a tapped delay line (TDL)-based wideband geometric model in \cite{Cheng2009geometry}, which is capable of investigating channel statistics for different time delays as well as the impact of vehicular traffic density.

\subsubsection{Non-geometric Stochastic Models}
Non-geometric stochastic models adopt a stochastic approach to model vehicular channels without assuming any underlying geometry \cite{Wang2009vehicle}. In particular, one of the most widely used non-geometric stochastic models is the TDL model \cite{Marum2007six}, which describes the channel by a finite impulse response filter with a number of taps, each associated with different delays and different types of Doppler spectra as well as different amplitude statistics (e.g., Ricean or Rayleigh). For example, the complex channel impulse response of a V2V channel can be modeled as the superposition of $L$ taps \cite{Wang2009vehicle,Marum2007six},
\begin{align}
h(\tau, t) = \sum\limits_{l=1}^L\left( \sum\limits_{n=1}^{N}\alpha_{l,n} e^{j2\pi f_{D,l,n}t} \right) \delta(\tau -\tau_l),
\end{align}
where each tap comprises $N$ unresolvable subpaths with amplitudes $\alpha_{l,n}$ and Doppler frequency as $f_{D,l,n} = vf_c\cos(\beta_{l,n})/c$, with $v$, $f_c$, $\beta_{l,n}$ and $c$ being the relative velocity, the carrier frequency, the aggregate phase angle of the $n$th subpath, and the speed of light, respectively.
Since each tap contains several unresolvable subpaths, almost arbitrary Doppler spectra can be synthesized in the TDL models.
The TDL models developed in \cite{Marum2007six} have been accepted and parameterized as the standard V2V channel model in IEEE 802.11p \cite{ieee2010ieee}.
They are based on the WSSUS assumption and have not considered the impact of non-stationarity of channel statistics, which can be incorporated by modeling multipath persistence via Markov chains as in \cite{Sen2008vehicle} or adding birth/death processes to account for sudden appearance/disappearance of a LoS component as in \cite{Wang2012improving}.


\section{Vehicular Channel Estimation}\label{sec:estimation}

Accurate and efficient channel estimation is a critical component in wireless communication systems.
It directly affects receiver design, e.g., channel equalization, demodulation, decoding, etc., as well as radio resource management for interference mitigation and performance optimization.
High Doppler spread in vehicular environments induces short channel coherence time, leading to challenges for vehicular channel estimation. We will focus on vehicular channel estimation in this section.

\begin{figure}[!t]
\centering
\includegraphics[width=0.46\textwidth]{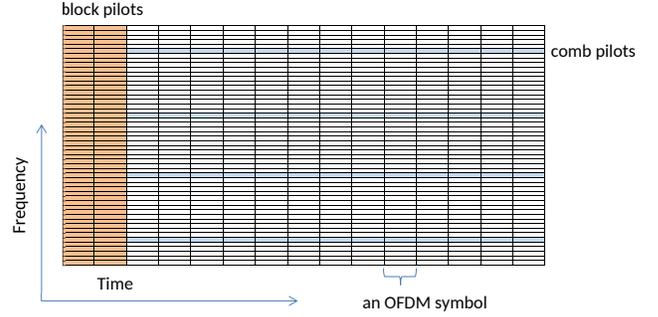}
\caption{Pilot arrangement in IEEE 802.11p \cite{ieee2010ieee,Bernado2010physical}.}\label{fig:pilot}
\end{figure}

In Fig.~\ref{fig:pilot}, we show the IEEE 802.11p training pilot arrangement \cite{ieee2010ieee,Bernado2010physical} to facilitate channel estimation for vehicular channels, which consists of two types of pilots: block and comb pilots. The least-square (LS) estimate \cite{Bernado2010physical,Beek1995channel} based on the block pilots is first performed to get the channel frequency response. The estimated channel is given by
\begin{align}
\mathbf{h}_{\text{ls}}=\mathbf{X}_b^{-1}\mathbf{y}_b,
\end{align}
where ${\mathbf X}_b$ contains the block training pilots and $\mathbf{y}_b$ is the corresponding received signal. Such LS estimation inevitably has performance limit due to fast channel variation and sparse placement of block pilots.
To remedy this problem, it has been proposed in \cite{Kim2008midamble} to periodically insert midambles pilots in the data field to provide channel update and tracking after the initial channel estimation based on the block pilots.
However, as noted in \cite{Lin2010novel}, such a midamble insertion scheme will lower the spectrum utilization efficiency due to the reduced ratio of data symbols in each frame.
In \cite{Lin2010novel}, a Zadoff-Chu sequence-based time domain LS estimation scheme has been developed to improve the estimation accuracy, where the pseudo noise sequence is inserted into the prefix to combat severe multipath fading without sacrificing spectrum efficiency.

It has been shown in a large body of literature \cite{Edfors1998ofdm,Li1998robust,Li2000pilot,Liu2014channel} that statistics of mobile radio channels, such as channel correlation over the time and frequency domains, can be exploited to improve channel estimation performance.
As a result, after performing initial LS estimate over the block pilots, the linear minimum mean-square-error (MSE) filtering \cite{Edfors1998ofdm} in the time domain over comb pilots has been implemented in \cite{Bernado2010physical} to track channel variation over time, yielding
\begin{align}\label{eq:lmmse}
\hat{\bf h}_{\text{lmmse}} = {\bf R_{h}}\left({\bf R_{h}} + \sigma^2(\mathbf{X}_c\mathbf{X}_c^H)^{-1}\right)^{-1}\mathbf{X}_c^{-1}\mathbf{y}_c,
\end{align}
where $\mathbf{R_h} = {\mathbb E}\{\mathbf{hh}^H\}$ is the time domain correlation matrix of channels, $\mathbf{X}_c$ is the comb training pilots, $\mathbf{y}_c$ is the received signal, and  $\sigma^2$ is the noise power.

For the WSSUS environments, the correlation of the channel frequency response at different time and frequency, $r_H(\Delta t, \Delta f)$, can be decoupled as the product of that in the time domain, $r_t(\Delta t)$, and that in the frequency domain, $r_f(\Delta f)$, i.e., $r_{H}(\Delta t, \Delta f) = r_t(\Delta t) r_f(\Delta f)$ \cite{Li1998robust}, where $r_t(\Delta t)$ depends on the vehicle speed or, equivalently, the Doppler shift, and $r_f(\Delta f)$ depends on multipath delay spread. As such, it can be separately estimated by averaging a large number of LS estimates over long time to get the frequency correlation and averaging over many subcarriers in orthogonal frequency division multiplexing (OFDM) systems to get the time domain correlation. For instance, the time domain correlation in \eqref{eq:lmmse} can be estimated by \cite{Bernado2010physical,Liu2014channel}
\begin{align}
\mathbf{R_h} \approx \frac{1}{K}\sum\limits_{k=1}^{K} \hat{\bf h}_{k} \hat{\bf h}_{k}^H, \nonumber
\end{align}
where $K$ is the number of comb pilot subcarriers and $\hat{\bf h}_{k}$ is the LS estimate of the channel at the comb pilot positions across all OFDM symbols between two block pilots.
Along this line, an approach using Wiener filter in the frequency domain has been proposed in \cite{Nuckelt2011performance} to search for the optimal coefficients to minimize the MSE of channel estimates, using the received signal-to-noise ratio (SNR) and channel statistics including maximum excess delay and the shape of the channel PDP.

Enhanced channel estimation techniques for IEEE 802.11p have been extensively studied over the past few years \cite{Zemen2012iterative,Zemen2012adaptive,Fernandez2012performance,Bourdoux2011channel,Zhao2013channel,Kim2014time}, which generally use decision-directed approaches to enhance the channel estimates.
In these schemes, the decided-data is further treated as pilots and the channel is re-estimated, which may even be performed iteratively.
A generalized discrete prolate spheroidal sequence based iterative approach has been developed in \cite{Zemen2012iterative}, where data detection and channel estimation are performed iteratively.
The number of iterations required has been further reduced by the adaptive estimation scheme proposed in \cite{Zemen2012adaptive}, which accounts for the non-stationarity of vehicular channels through adapting the robust reduced-rank Wiener filter to varying propagation conditions on a frame-by-frame basis.
In \cite{Fernandez2012performance}, the decision-directed channel estimate has been enhanced through averaging in both the time and the frequency domains to reduce the effects of noise and erroneous channel estimates based on unreliable data subcarriers.
Similarly, additional smoothing has been proposed in \cite{Bourdoux2011channel} to alleviate the problem of error propagation in the low SNR regime.
The constructed data pilot (CDP) scheme in \cite{Zhao2013channel} has utilized the data symbols to form pilots and then exploited the channel correlation between two consecutive symbols to improve estimation accuracy.
The scheme in \cite{Kim2014time} recovers channel estimates at unreliable data subcarriers based on frequency domain interpolation of the reliable data pilots, which are selected through the reliability test on the initial channel estimates.


For vehicular communications, the V2I and V2V channels typically exhibit sparse scattering, such that the delay-Doppler spectrum is sparse \cite{Paier2010vehicular}.
Such channel sparsity has been exploited in \cite{Beygi2017structured,Beygi2015nested} by representing the low-rank structure as a summation of a few rank-one atoms to improve channel estimation performance while taking into account the channel leakage caused by finite block length and transmission bandwidth.
For cellular-assisted V2X systems, the low density of pilot symbols in the current long term evolution (LTE) standards sometimes leads to poor accuracy of channel estimation \cite{Schwarz2017signal}.
As a remedy, four demodulation reference signals (DMRS), rather than just two normally utilized in LTE standards, have been suggested in \cite{Sun2016support} to augment estimation accuracy.
In addition, an adaptive pilot pattern design scheme has been proposed in \cite{Simko2013adaptive} to follow changing channel statistics and adapt to the Doppler and delay spread of vehicular channels, which has been shown to outperform the LTE standard compliant schemes.

%
\section{Modulation for High Mobility Channels}\label{sec:modulation}

Due to the high Doppler spread caused by vehicle mobility, the multicarrier modulation scheme is more susceptible to ICI in vehicular communications.
As a result, modulation and equalization designs with special consideration of high mobility are necessary.
In this section, we review major research findings that target ICI cancellation for OFDM systems in time-varying vehicular channels and discuss some new waveform designs developed under the framework of the fifth generation cellular technology (5G) that can potentially improve vehicular system performance.

\subsection{ICI Analysis for OFDM}
It has been shown \cite{Russell1995interchannel,Robertson1999effects,Li2001bounds} that time variation within an OFDM symbol destroys orthogonality among subcarriers and and introduces ICI. If not properly accounted for, ICI would result in an error floor, which increases with vehicle mobility and carrier frequency.

To understand ICI in OFDM systems, we consider an OFDM signal in the time domain as
\begin{align}\label{eq:ofdm}
s(t) = \sum\limits_{k}s[k]e^{j2\pi f_k t}, 0 \le t \le T_s,
\end{align}
where $f_k = f_0 + k\Delta f$ is the frequency of the $k$th subcarrier with $\Delta f = 1/T_s$ being the subcarrier spacing and $T_s$ being the OFDM symbol duration.
After passing the OFDM signal in \eqref{eq:ofdm} through the time-varying wireless channel with an impulse response, $h(t,\tau)$, as given in \eqref{eq:impulse}, the demodulated signal at the $m$th subcarrier (without considering channel noise) will be
\begin{align}
\tilde{s}[m] = a_0 s[m] + \sum\limits_{k\ne m}a_{m-k}s[k],
\end{align}
where $a_0$ represents the attenuation and phase shift of the desired signal and $a_l$'s with $l\ne 0$ denote the ICI terms.
They depend on the path delays and time-varying path attenuation.

To gain further understanding about how severe the ICI is, we define the ICI power as
\begin{align}
P_{\text{ICI}} = \mathbb{E}\left[\left|\sum\limits_{l\ne 0}a_l s[m-l]\right|^2 \right].
\end{align}
The exact expressions for ICI power have been derived in \cite{Russell1995interchannel,Robertson1999effects} in a closed form for different Doppler spectra.
Moreover, tight upper and lower bounds on ICI power have been derived in \cite{Li2001bounds}.
A universal (tight) upper bound, dependent only on the maximum Doppler shift, $f_d$, and the symbol duration, $T_s$, has been further derived as
\begin{align}
P_{\text{ICI}} \le \frac{\pi^2}{3}(f_d T_s)^2.
\end{align}
Based on the derived bound, the effect of Doppler spread can be reduced or made negligible compared with other impairments if the OFDM symbol duration $T_s$ is chosen such that $f_d T_s$ is small.

\subsection{Mitigating ICI for OFDM}

To reduce the effect of ICI, several methods have been proposed to estimate the ICI and then perform ICI cancellation, sometimes in an iterative fashion.
A frequency-domain equalization scheme has been developed in \cite{Jeon1999equalization} that assumes the channel impulse response varies linearly during an OFDM block and has shown desirable performance under low Doppler and delay spread conditions.
The ICI cancellation performance in high Doppler and delay spread channels has been further enhanced in \cite{Mostofi2005ICI}, where two ICI mitigation methods have been proposed based on piece-wise linear approximation for channel temporal variation.
A polynomial representation for multipath channel variation has been assumed in \cite{Hijazi2009polynomial}, where the channel gains have been estimated using averaged LS estimates and the ICI has been reduced by successive interference suppression, in an iterative manner.
In \cite{Tang2007pilot}, the basis expansion model has been introduced to approximate time-varying channels in OFDM systems, which has reduced the number of unknowns to be estimated and improved system performance.
In \cite{Nissel2015doubly}, an MMSE channel estimation and ICI mitigation scheme has been proposed, where the channel estimation accuracy and equalization performance have been improved through the tree-step iterative ICI cancellation technique.

As an alternative to pilot-assisted ICI cancellation, which may reduce spectral efficiency, there have been works to design OFDM transmission with self-ICI-cancellation capability.
In \cite{Zhao1998intercarrier}, the partial correlative coding (PRC) in the frequency domain has been proposed to mitigate ICI in OFDM systems.
A general frequency-domain PRC has been further studied in \cite{Zhang2003optimum}, where the optimum weights for PRC to minimize ICI power has been derived.
Wiener filtering in the downlink and transmit preprocessing in the uplink have been proposed in \cite{Ma2011statistics} to exploit correlation between the desired signal and the aggregate ICI in presence of a deterministic LoS path.
An ICI self-cancellation scheme, which transmits each symbol over a pair of adjacent subcarriers with a $180^{\circ}$ phase shift, has been developed in \cite{Zhao2001intercarrier} to suppress ICI at the expense of a reduced transmission rate.
This has been extended in \cite{Armstrong1999analysis} by mapping data onto larger groups of subcarriers such that higher order ICI cancellation can be achieved.
In \cite{Chang2007novel}, another ICI self-cancellation scheme with flexible pre- and post-processor designs has been further proposed to achieve a trade-off between ICI mitigation and sum throughput.
Based on \cite{Zhao2001intercarrier}  and \cite{Chang2007novel}, a general reduced-rate OFDM transmission scheme for ICI self-cancellation has been developed in \cite{Ma2012reduced}, which cancels the ICI through optimized transmit and receive processing designed with channel statistics only.

\subsection{New Waveform Design}

Recently, under the framework of 5G, a wide variety of new waveform designs have also been investigated to strike a balance between residual ICI or intersymbol interference (ISI) and achieved spectral efficiency \cite{Banelli2014modulation}.
Some of the waveform parameters can even be adapted to the wireless channel statistics, such as Doppler and delay spread.
For instance, subcarrier filtering-based filter bank multicarrier (FBMC) \cite{Farhang2011ofdm} and generalized frequency division multiplexing (GFDM) \cite{Fettweis2009gfdm}, and subband filtering-based universal filtered multicarrier (UFMC) \cite{Vakilian2013ufmc} have the required flexibility to support waveform parameters adaption over subcarriers and subbands. They may provide the basis for channel-adaptive modulation in vehicular communications.
Alternatively, the transmit data can be modulated in the delay-Doppler domain as proposed by the orthogonal time frequency and space (OTFS) modulation developed in \cite{monk2016otfs}, where channels with high Doppler shifts can be expressed in a stable model, thus reducing the costs of tracking time-varying vehicular channels.

To manage the dramatically increased number of simultaneous connections due to high vehicle density, the non-orthogonal multiple access (NOMA) scheme \cite{Saito2013noma,Ding2016application} can be used in vehicular communications.
Different from traditional orthogonal multiple access schemes, NOMA allows each resource block (e.g., time/frequency/code) to be employed by multiple users at the same time and therefore supports more connectivity at the cost of extra interference and detection complexity at the receiver side.
Since sophisticated hardware and ample power supply can be expected in vehicles, the implementation complexity of NOMA may not be of an issue in vehicular communications.

\section{Resource Allocation}\label{sec:resrouce}
Resource allocation plays a key role in mitigating interference and optimizing resource utilization for vehicular communications. Though it is most often regarded as a MAC layer issue, we acknowledge its close interaction with the physical layer design \cite{Ge2009phy} and review some latest treatment of this topic in this section. In particular, we will discuss the state-of-the-art techniques on resource allocation design in cellular-assisted V2X communications based on D2D networks due to its significant potential to improve vehicular communications performance \cite{Sun2016radio,Araniti2013lte}.

The D2D communications enable direct transmission between devices in proximity without routing through the base station and have been the subject of much recent research \cite{Asadi2014survey,Feng2013device}.
Resource allocation is a critical issue in enabling D2D underlay in cellular networks to improve spectrum utilization efficiency and reduce transmission latency.
For example, a three-step approach has been proposed in {\cite{Feng2013device}} to design power control and spectrum allocation to maximize system throughput with a minimum signal-to-interference-plus-noise ratio (SINR) guarantee for both cellular and D2D links.
In {\cite{Janis2009interference}}, D2D transmit powers have been regulated by the base station such that the SINR of D2D links is maximized while the interference experienced by the cellular links is kept at an acceptable level.

Vehicular channels experience fast temporal variation due to vehicle mobility \cite{Zhao2013channel}.
Therefore, traditional resource allocation designs for D2D communications with full CSI are no longer applicable due to the formidable signaling overhead to track channel variation on such a short time scale.
Applying D2D techniques to support vehicular communications thus mandates further study on radio resource management accounting for fast vehicular channel variation.
Along this line, a feasibility study of D2D for vehicular communications has been performed in \cite{Cheng2015D2D} to evaluate the applicability of D2D underlay in supporting joint V2V and V2I connections in cellular networks.
It has been shown in \cite{Cheng2015D2D} that D2D-aided vehicular communications can outperform the traditional V2V-only mode, the V2I-only mode, or the V2V overlay mode in terms of achievable transmission rates.

In \cite{Botsov2014location}, a heuristic location dependent uplink resource allocation scheme has been proposed for D2D terminals in vehicular networks, which features spatial resource reuse with no explicit requirement on full CSI and, as a result, significantly reduces signaling overhead.
A framework comprising vehicle grouping, reuse channel selection, and power control has been developed in \cite{Ren2015power} to maximize the sum rate or minimally achievable rate of V2V links while restraining the aggregate interference to the uplink cellular transmission. A series of simplifications have been applied to the power control problem to reduce the requirement of full CSI and the dependence on centralized control as well as the computational complexity.
In \cite{Sun2016radio}, latency and reliability requirements of V2V communications have been transformed into optimization constraints computable using large-scale fading information only.
A heuristic algorithm has been developed to address the proposed radio resource management optimization problem, which adapts to the large-scale fading of vehicular channels, i.e., pathloss and shadowing that vary on a slow time scale.
Similar system setups have been further considered in \cite{Sun2016cluster}, where multiple resource blocks are allowed to be shared not only between cellular and D2D users but also among different D2D-capable vehicles.
In \cite{Cheng2016performance}, power control based on channel inversion using pathloss information and D2D mode selection based on biased channel quality have been proposed to enable vehicular D2D communications (V-D2D) in cellular networks. Two representative performance metrics, SINR outage probability and network throughput, have been analyzed in the established theoretical framework.

Different from the existing works, we have taken a joint consideration of the heterogeneous QoS requirements of both V2I and V2V links, i.e., large capacity for bandwidth intensive V2I links and high reliability for safety-critical V2V links, in \cite{Liang2017resource,Liang2017spectrum}.
We maximize the sum V2I ergodic capacity with V2V reliability guaranteed, which requires only large-scale fading information and thus reduces network signaling overhead to track rapidly changing vehicular channels. Meanwhile, the fast fading effects have been rigorously treated. Accordingly, the resource allocation problem for vehicular communications can be formulated as \cite{Liang2017resource}
\begin{align}
 &\max\limits_{\overset{\{\rho_{m,k}\}} {\{P_m^c\}, \{P_k^d\}}} ~\sum_{m}{\mathbb E}\left[\log_2\left(1 + \gamma_{m}^c\right) \right] \label{eq:optV2V}\\
 & ~~~~~ \text{s.t.} ~~~~~~ {\mathbb E}\left[ \log_2(1 + \gamma_m^c)\right] \ge r_0^c,  \forall m  \tag{\ref{eq:optV2V}a}\\
 & ~~~~~~~~~~~~~~\text{Pr}\{\gamma_k^d \le \gamma_{0}^d \} \le p_0, \forall k,  \tag{\ref{eq:optV2V}b}
\end{align}
where $P_m^c$ and $P_k^d$ denote the transmit power levels of the $m$th V2I and the $k$th V2V transmitters, respectively, $\gamma_m^c$ and $\gamma_k^d$ are the received SINRs, $\rho_{m,k}\in\{0,1\}$ is the spectrum reuse indicator with $\rho_{m,k}=1$ implying the $k$th V2V link reuses the spectrum of the $m$th V2I link (whose spectrum has been orthogonally allocated) and $\rho_{m,k}=0$ otherwise, $r_0^c$ is the minimum V2I capacity guarantee, $\gamma_0^d$ is the minimum SINR threshold required to establish a reliable V2V link, and $p_0$ is the tolerable V2V outage probability that is typically very small. The global optimum to the optimization problem has been found in \cite{Liang2017resource} with a low-complexity algorithm employing graph theoretic and standard optimization tools.


\section{mmWave for Vehicular Communications}\label{sec:mmWave}

\begin{figure}
\centering
\subfigure[]{\includegraphics[width=0.33\textwidth]{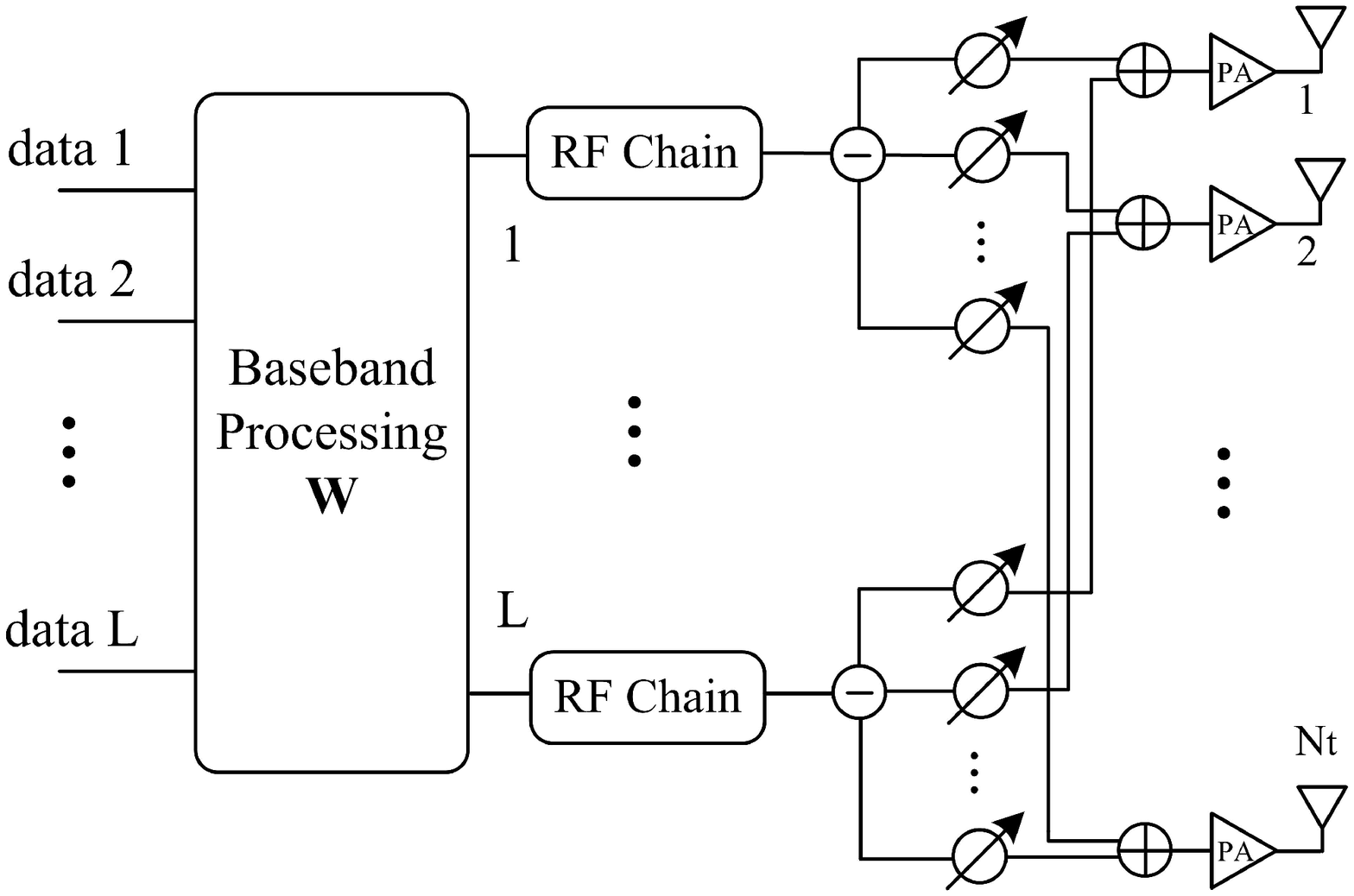}}
\subfigure[]{\includegraphics[width=0.31\textwidth]{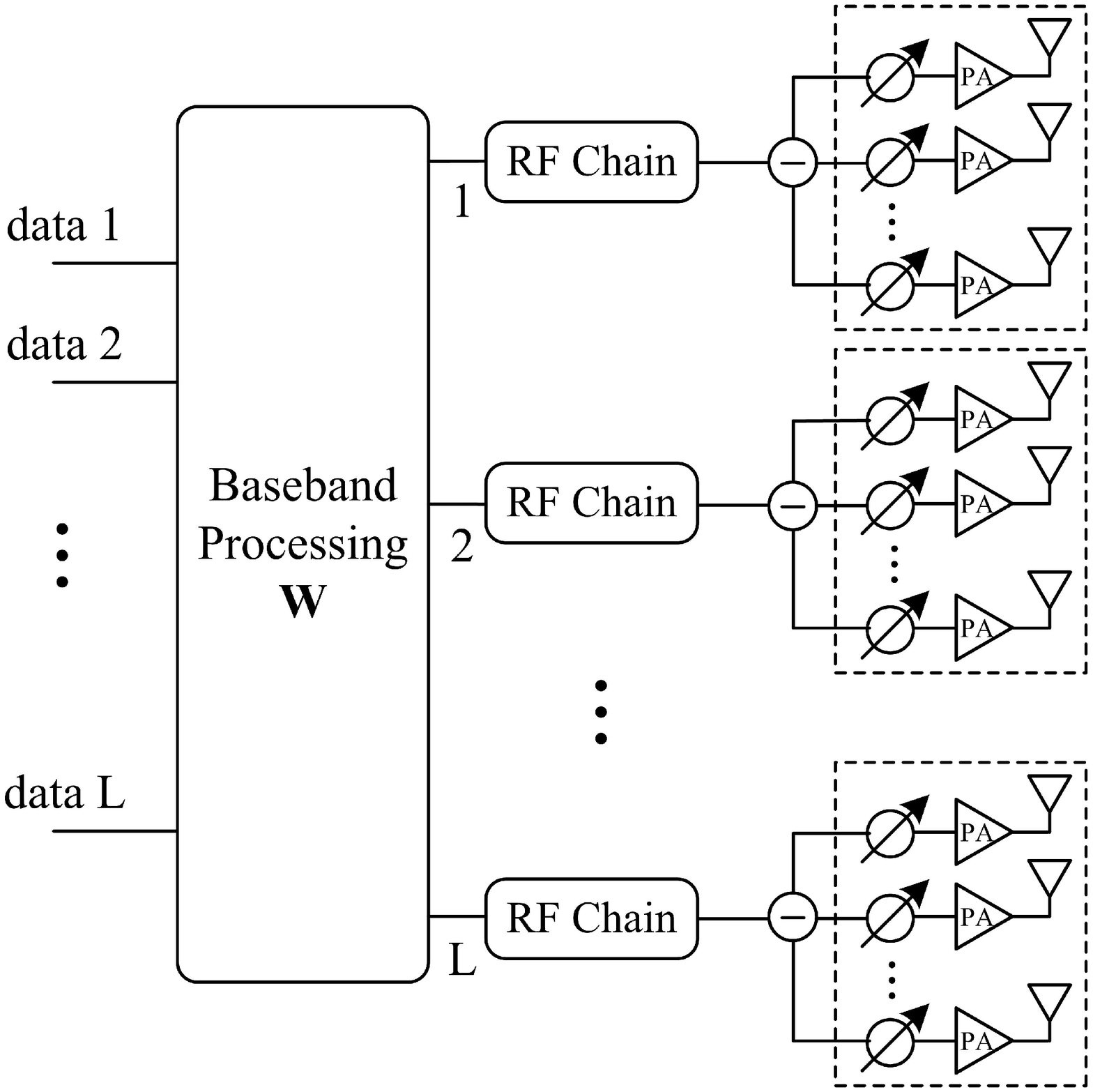}}
\caption{Two hybrid processing structures for mmWave communications: a) fully-connected; b) array-of-subarray.}\label{fig:hybrid}
\end{figure}

The mmWave band, i.e., 30-300 GHz, is desirable for vehicular communications due to its significant potential to improve system performance with order of magnitude larger bandwidth.
Thanks to the small wavelength at the mmWave frequencies, an excessive number of antennas can be packed in a small form factor and generate narrow directional beams to compensate for the severe mmWave propagation loss.
With directional transmission and narrower beams, Doppler spread at mmWave frequencies can be effectively managed even in high mobility environments \cite{Va2016impact}.
Thus, it is intriguing to take advantage of the mmWave band to enhance the performance of vehicular communications.

\subsection{Hybrid Processing Architecture}
The high cost of mmWave hardware and the large number of antenna elements preclude the possibility to support each antenna element with a dedicated radio frequency (RF) chain in mmWave systems \cite{Heath2016overview,Roh2014mmWave,Liang2014how}.
To alleviate this issue, mmWave communications can rely on analog RF processing using phase shifters combined with low dimensional baseband processing to achieve MIMO benefits, including multiplexing and array power gains. Such a hybrid analog and digital processing structure has gained considerable attention \cite{Ayach2014spatially,Zhang2015massive}.
Generally, there are two possible architectures for hybrid processing as shown in Fig.~\ref{fig:hybrid}.
The first one connects each RF chain output with all antenna elements through a phase-shifting network \cite{Ayach2014spatially,liang2014low} as shown in Fig.~\ref{fig:hybrid}(a).
Or, only a subset of the antenna elements is connected to each RF output via a phased array in the second one \cite{Lin2016terahertz} as shown in Fig.~\ref{fig:hybrid}(b). The second architecture reduces the implementation complexity at the cost of discounted processing flexibility compared with the first one. Interested readers can refer to \cite{Lin2016terahertz,Han2015large} for detailed comparison and analysis of two architectures in terms of energy and spectral efficiency.


\subsection{mmWave for Vehicular Communications}
The mmWave band exhibits several distinctive propagation characteristics compared with its low frequency counterparts including, e.g., vulnerability to blockage, extra atmospheric absorption, and sparse scattering \cite{Heath2016overview,Va2016millimeter}.
There have been some research efforts devoted to understanding mmWave channels for vehicular communications through measurements and modeling \cite{Wyne2011beamforming,Dor2011millimeter,Sato2001propagation,Schneider2000impact,Karasawa2002multipath,Va2016millimeter}.
Due to higher carrier frequency and rapid vehicle movement, severe Doppler effects are among the most critical concerns in exploiting mmWave bands for vehicular communications.
However, as reported in \cite{Va2016impact}, such effects can be substantially reduced through directional beamforming, where the optimal beamwidth has been derived considering the pointing error due to mobility.
It has been shown that the beam coherence time for mmWave vehicular communications with directionally aligned beams, which is an effective measure of beam realignment frequency, can be much larger than the channel coherence time.
This implies beam alignment in every beam coherence period outperforms beam alignment in every channel coherence time when overheads are taken into account.

Location-aided angle-of-departure (AoD)/angle-of-arrival (AoA) estimation for mmWave vehicular communications has been proposed in \cite{Garcia2016location} based on beamforming codebooks, which has been shown to significantly reduce channel estimation time and speed up initial access.
Out-of-band measurements at sub-6 GHz have been utilized in \cite{Ali2016estimating} to assist estimating mmWave vehicular channels through transforming the spatial correlation matrix at the lower frequency to mmWave bands. Two different methods, a non-parametric approach based on interpolation/extrapolation and a parametric approach based on estimates of AoA and angular spread, have been proposed to perform the correlation matrix translation.
A beam switching scheme leveraging the position information of the train has been proposed in \cite{Va2015beamSwitching}, where traditional beam sweeping has been shown to be inefficient because the use of quasi-omni beams will result in large Doppler spread and reduce the channel coherence time.
In \cite{Va2016beam}, beam design has been further optimized in a mmWave V2I scenario, where beam switching based on position prediction is employed. The performance of equal beamwidth design and equal coverage design has been compared in terms of average rates and outage probabilities, which provides insights into the design of beamwidth and the allowable beam overlapping.

Most existing works for mmWave vehicular communications rely on a single-phased array
\cite{Va2016impact,Va2016beam,Garcia2016location}. As beam sweeping may be inefficient for mmWave vehicular communications due to high mobility, beam tracking and switching based on vehicle motion and location prediction are widely adopted. However, inaccuracy of the vehicle location prediction and the accumulation of the phased-array switching time for different beam directions will lead to severe beam misalignments and pointing error. As a result, several beam realignments are required during the transmission, which will introduce extra overhead. To deal with these issues, multiple antenna arrays, such as array of subarrays \cite{Zhang2015massive,Lin2016terahertz}, shall be further investigated for mmWave vehicular communications, which can potentially achieve much better performance than the single-phased arrays.


\section{Challenges and Opportunities}\label{sec:challenges}
To enable highly reliable and efficient vehicular communications, there are still many challenges that need to be investigated and addressed.
They also bring abundant research opportunities to both academia and industry.
Some of these challenges and opportunities are identified and discussed in this section.

\subsection{Channel Measurement and Modeling}

There have been lots of research activities on the measurement and modeling of V2V channels, typically at the DSRC bands, i.e., 5.9 GHz under the umbrella of IEEE 802.11p \cite{Mecklenbrauker2011vehicular}.
It has been understood from these efforts that V2V channels exhibit higher Doppler spread and possible non-stationarity of channel statistics, making such channels more challenging to work with when designing vehicular transmission schemes.
Most of these works have focused on historically common propagation scenarios, e.g., urban, suburban, rural, highway.
It is thus necessary to extend such measurement and modeling efforts to many other environments that are also vital for vehicular communications, especially when safety-critical applications are to be supported, including tunnels, bridges, parking lots, to name a few \cite{Viriyasitavat2015vehicular}.
In addition, although V2I channels are commonly regarded as similar to conventional cellular channels, where one end is stationary while the other is moving, they are unique in many aspects.
Some obvious examples include the impact of vehicle antenna placement on signal propagation characteristics, the proximity of V2I roadside units to the road, and typically lower heights of V2I roadside unit antennas compared with cellular base station antennas.
All these necessitate further investigation into V2X channels, not to mention the sparsity of literature on modeling vehicle-to-pedestrian (V2P) channels.

In view of the advent of cellular-assisted V2X communications \cite{Seo2016LTE,Schwarz2017signal}, more works are expected to gain deeper understanding about vehicular channels at cellular bands, in addition to DSRC spectrum.
Meanwhile, the industry is faced with increasing demands and popularity of exploiting mmWave bands for high rate data exchange among vehicles \cite{Choi2016millimeter}. However, channel measurement and modeling of vehicular channels at mmWave frequencies are very limited. Significantly more research efforts are needed in this regard, studying the impacts of moving scatters, shadowing by other vehicles, non-stationarity, etc. on vehicular channel propagation characteristics.

\subsection{Resource Allocation for Vehicular Service Heterogeneity}

To meet the heterogeneous QoS requirements of vehicular networks, proper radio resource management with vehicle mobility consideration is a key enabler.
While the IEEE 802.11p-based DSRC and ITS-G5 standards have established the foundation for vehicular communications, the cellular-assisted V2X communication technology is expected to be more powerful and efficient \cite{Schwarz2017signal,Sun2016support,Seo2016LTE}.
The spectrum available for its usage will expand from the traditional cellular bands to the DSRC spectrum, with the potential of including even the much higher mmWave frequencies.
How to coordinate reliable vehicular transmission over such a wide variety of spectrum is a critical challenge and therefore calls for novel resource allocation schemes that can effectively manage spectrum heterogeneity (e.g., game theoretic spectrum sharing approach \cite{Zhou2016toward}) to be developed.

The D2D communications, as an indispensable part of the cellular-assisted V2X systems, are expected to play an important role in V2V direct connections.
Spectrum reuse between D2D links and cellular links is normally a vital technique to improve spectrum utilization efficiency \cite{Feng2013device}, which necessitates the development of efficient interference control.
Different from the current standardized D2D technique specialized in providing public safety, the next generation D2D-enabled V2X communications will support advanced safety services, such as higher bandwidth raw sensor data sharing \cite{Choi2016millimeter}.
Besides, rapid temporal variation of vehicular channels challenges traditional resource allocation schemes for D2D-based networks due to its significant signaling overhead to acquire CSI at the central controller. For instance, channel inaccuracy due to estimation error and feedback delay \cite{Awad2009ergodic} is expected to be more severe in rapidly changing vehicular environments.
Therefore, novel resource allocation designs tailored for D2D-based vehicular networks with rigorous treatment of channel variation and vehicular service differentiation are desired.

\subsection{mmWave-Enabled Vehicular Communications}

Exploiting the mmWave bands for vehicular communications is an attractive option to support advanced safety and information and entertainment (infotainment) services in vehicular networks due to the abundance of spectrum availability \cite{Choi2016millimeter}. That said, works on measurement and modeling of wireless propagation channels for vehicular communications at mmWave frequencies are still very limited.
More research efforts are thus needed to better understand unique characteristics of mmWave vehicular channels, including higher attenuation due to gaseous and rain absorption, richer scattering and reflection due to reduced wavelength, poorer diffraction and penetration, and higher susceptibility to signal blockage from moving vehicles or pedestrians.
Other impact factors of vehicular channels, such as antenna array positions, the sizes and types of vehicles, and channel statistics variation due to fixed and moving scatterers, also need to be studied and properly modeled at the mmWave frequencies.

Another significant challenge for enabling vehicular communications over mmWave spectrum is the potentially large overhead for establishing reliable mmWave links, which is made even worse by the constant moving of vehicles and the complex vehicular channels at mmWave frequencies. Initial results from \cite{Va2015beamSwitching,Va2016beam} have demonstrated that beam tracking designs that adapt mmWave beam directions to predicted vehicle locations represent a promising solution. The majority of these works deal with LoS propagation, which is a reasonable assumption in open environments, such as suburban and highways.
However, in dense urban environments, non-line-of-sight (NLoS) propagation is usually the case, which brings challenges to the beam tracking algorithms due to the potentially unbounded location prediction error as well as severe beam misalignment. More efforts are thus needed to address such issues. This problem also closely interacts with the investigation of the coverage of mmWave networks \mbox{\cite{Bai2014coverage}} since it determines the frequency of cell handover and the associated beam management overhead. It would be particularly interesting to see how the beam management scales with respect to the vehicle velocity as well as the ensuing measurement and processing complexity in mmWave environments.

\subsection{Vehicular Communications in 5G}
The 5G technology has spurred heated discussion in recent years \mbox{\cite{Andrews2014what}}, and three representative service categories, i.e., enhanced mobile broadband (eMBB), massive machine-type communication (mMTC), and ultra-reliable and low latency communication (uRLLC), have been defined by the international telecommunication union (ITU) in 2015 \mbox{\cite{IMT2015}}.
The 3GPP is now developing an OFDM-based 5G new radio (NR) as a unified air interface to support such a wide variety of device types, spectrum, and services.
In the meantime, V2X communications have been under active development as part of the 5G ecosystem. Potential 5G V2X use cases have been envisioned to include platooning, advanced driving, remote driving, etc \mbox{\cite{3GPPr15v2x22886}}.
There is no doubt that 5G V2X will bring new capabilities for connected vehicles and enable safer and more efficient driving in the near future.
In particular, 5G V2X will accelerate the fusion and sharing of various onboard sensor data, including, e.g., radar, LIDAR (light detection and ranging), and cameras, and facilitate accurate positioning as well as high definition local map rendering to help realize connected autonomous driving.

The unique characteristics of V2X communications complicate the 5G design in that the service requirements can easily span the three defined categories while demanding a unified treatment. Specifically, the critical nature of safety related V2X messages places stringent requirement on the link reliability and transmission latency, real-time sharing of rich onboard sensor data demands high communication throughput, and the sheer high density of vehicles on the road necessitates a large number of simultaneous connections.
These seemingly contradicting requirements challenge various physical layer designs. For example, in terms of coding schemes for V2X, the low latency transmission prefers the use of short codes while on the other end of the spectrum, high link reliability requires extremely low error rates where long codes appear a better fit. Among the coding candidates for 5G, i.e., Turbo, LDPC, and Polar coding, recent studies \mbox{\cite{Sybis2017channel}} show that none of them can do a perfect job in satisfying such diverse needs, where polar and LDPC codes outperform turbo codes for short block sizes, while the opposite is true for medium block sizes.
The problem is further complicated by the challenging wireless environment and the increasing Doppler effect.
Therefore, significant innovations, originating from virtually all aspects of the communications technology, such as disruptive network topology, novel modulation/coding schemes, and non-traditional estimation and detection/decoding designs, are expected to be developed to unleash the full potentials of the 5G V2X communications.

\section{Conclusions}\label{sec:conclusion}
In this article, we have provided a comprehensive overview of vehicular communications from the physical layer perspective.
Efficient and reliable vehicular communication systems will bring unprecedented benefits to the society while at the same also pose unparalleled challenges due to its unique characteristics.
Increase in terminal mobility and demanding QoS requirements, e.g., ultra low latency, high reliability, service heterogeneity, for vehicular networks render traditional wireless communications design inefficient if not ineffective.
A wide spectrum of research efforts attempting to alleviate such challenges have been reviewed, including development of novel channel models for vehicular environments, design of efficient channel estimation and modulation schemes accounting for high vehicle mobility, radio resource allocation designs, and exploiting mmWave bands for vehicular transmission.
We have further identified major under-explored issues and pointed out areas that require more attention.



\bibliographystyle{IEEEtran}
\bibliography{Refv2xSurvey}

\begin{IEEEbiography}[{\includegraphics[width=1in,height=1.25in,clip,keepaspectratio]{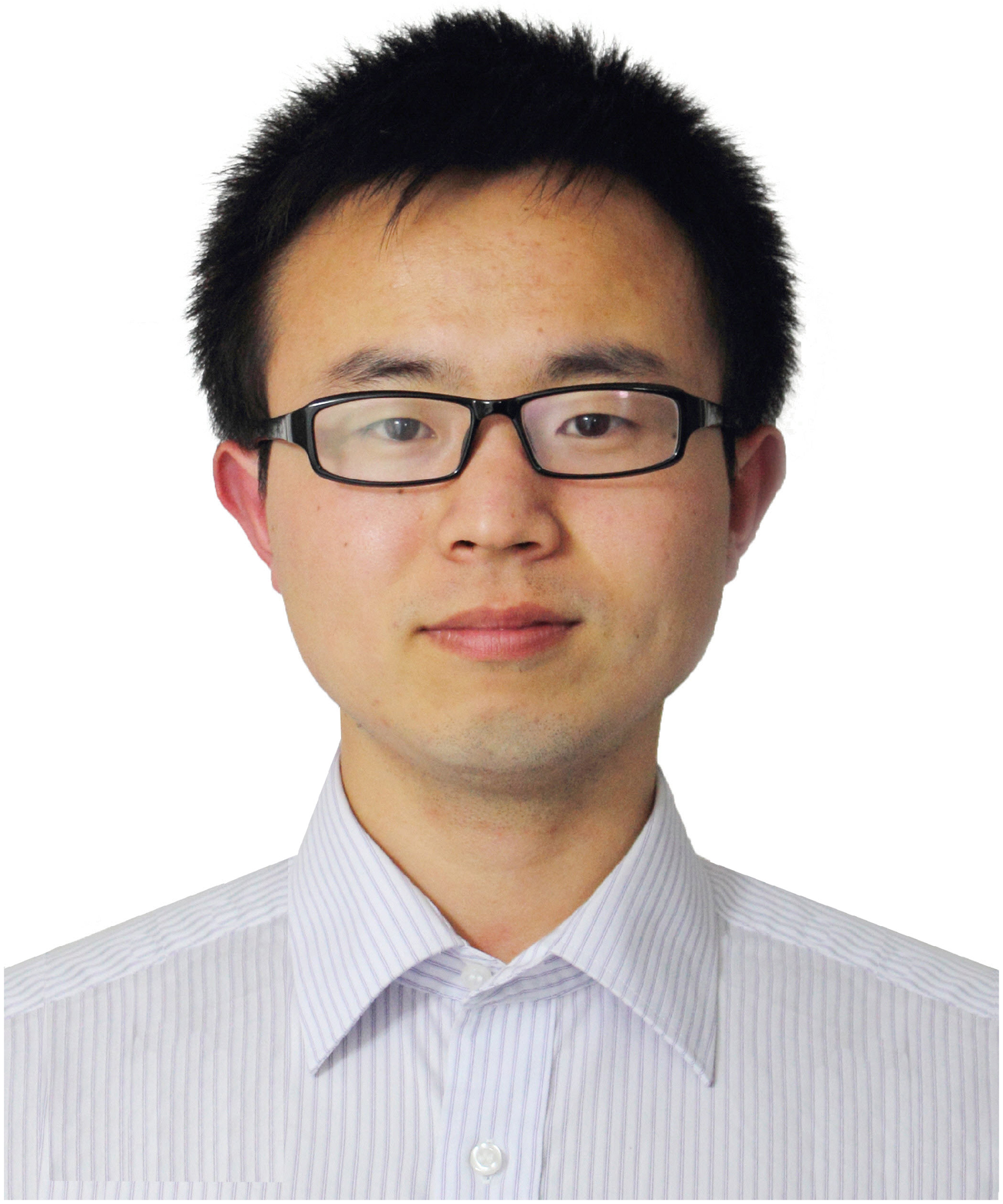}}]{Le Liang}
(S'12) received the B.E. degree in information engineering from Southeast University, Nanjing, China in 2012 and the M.A.Sc degree in electrical engineering from the University of Victoria, Victoria, BC, Canada in 2015. He is currently working towards the Ph.D. degree in electrical engineering at the Georgia Institute of Technology, Atlanta, GA, USA.
His research interests include vehicular communications, radio resource management, and MIMO systems.
\end{IEEEbiography}

\begin{IEEEbiography}[{\includegraphics[width=1in,height=1.25in,clip,keepaspectratio]{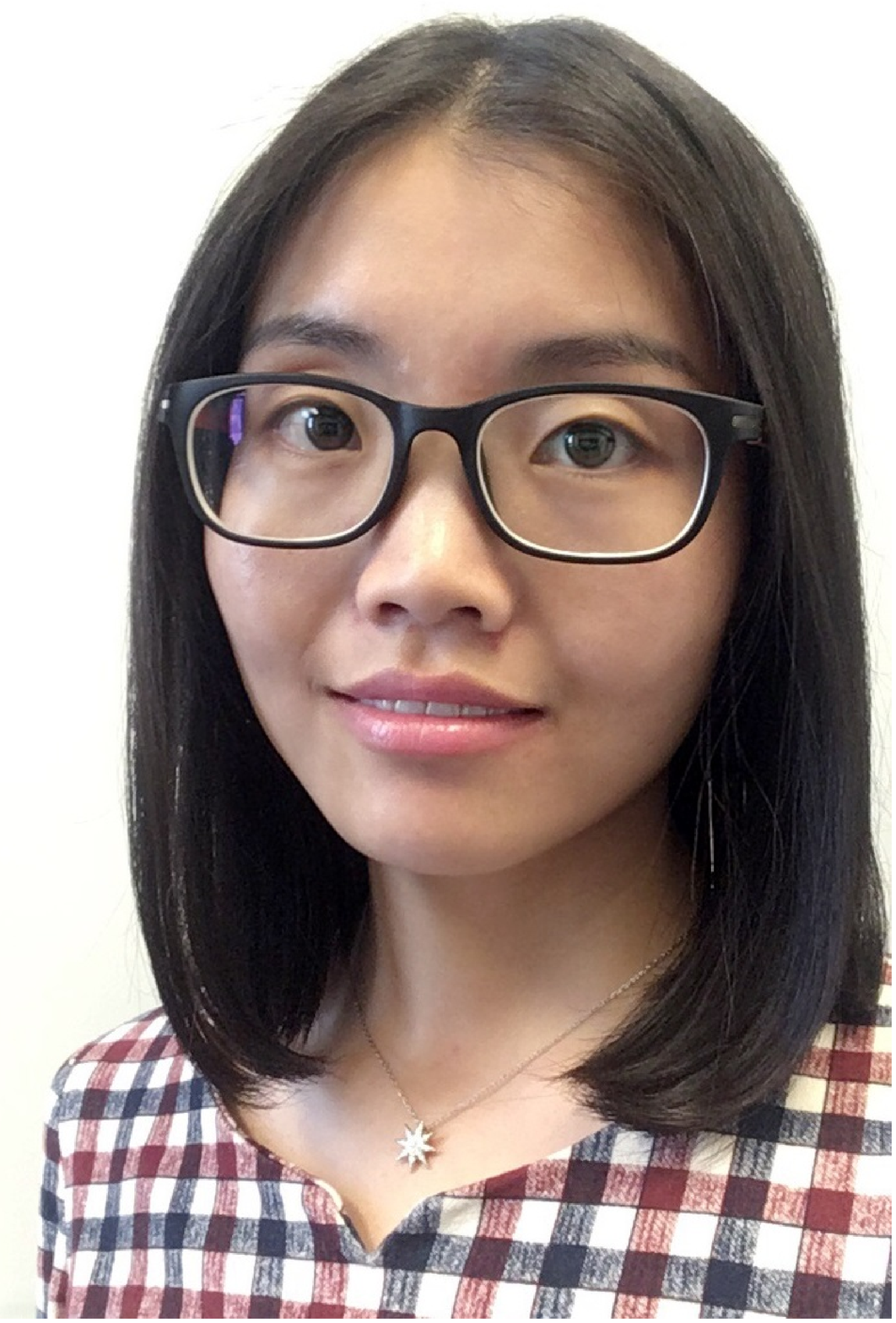}}]{Haixia Peng}
(S'15) received her M.S. and Ph.D. degrees in electronics and communication engineering and computer science from Northeastern University, Shenyang, China, in 2013 and 2017, respectively. She is currently a Ph.D student in the Department of Electrical and Computer Engineering at University of Waterloo, Canada. Her current research focuses on vehicular networks and autonomous vehicle communications. She served as a TPC member in {IEEE VTC-fall 2016\&2017, IEEE Globecom 2016\&2017, IEEE ICC 2017\&2018} conferences.
\end{IEEEbiography}

\begin{IEEEbiography}[{\includegraphics[width=1in,height=1.25in,clip,keepaspectratio]{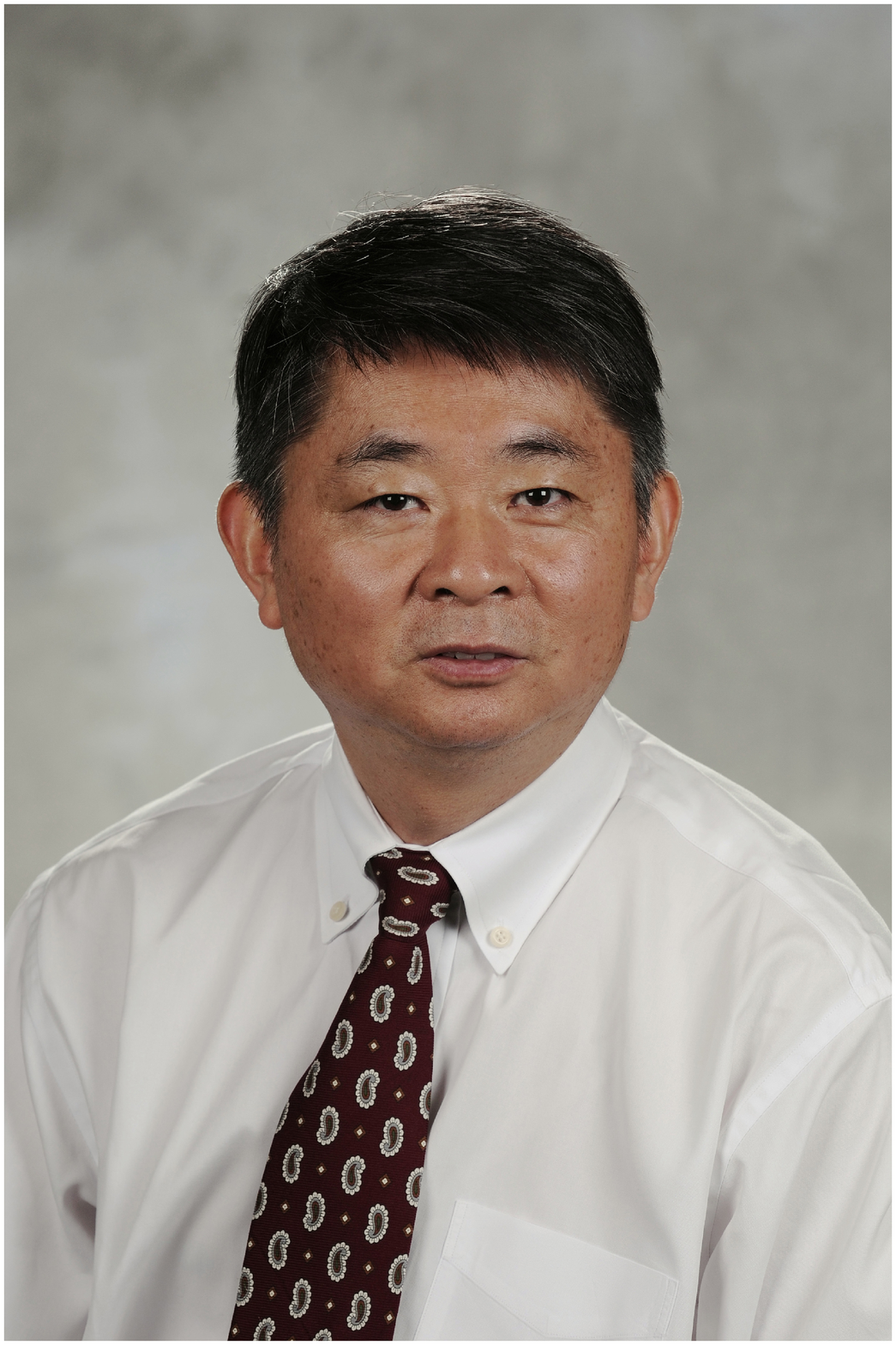}}]{Geoffrey Ye Li}
(S'93-M'95-SM'97-F'06) received his B.S.E. and M.S.E. degrees in 1983 and 1986, respectively, from the Department of Wireless Engineering, Nanjing Institute of Technology, Nanjing, China, and his Ph.D. degree in 1994 from the Department of Electrical Engineering, Auburn University, Alabama.

He was a Teaching Assistant and then a Lecturer with Southeast University, Nanjing, China, from 1986 to 1991, a Research and Teaching Assistant with Auburn University, Alabama, from 1991 to 1994, and a Post-Doctoral Research Associate with the University of Maryland at College Park, Maryland, from 1994 to 1996. He was with AT\&T Labs - Research at Red Bank, New Jersey, as a Senior and then a Principal Technical Staff Member from 1996 to 2000. Since 2000, he has been with the School of Electrical and Computer Engineering at the Georgia Institute of Technology as an Associate Professor and then a Full Professor. He is also holding a Cheung Kong Scholar title at the University of Electronic Science and Technology of China since 2006.

His general research interests include statistical signal processing and communications, with emphasis on cross-layer optimization for spectral- and energy-efficient networks, cognitive radios and opportunistic spectrum access, and practical issues in LTE systems. In these areas, he has published around 200 journal papers in addition to over 40 granted patents and numerous conference papers. His publications have been cited over 29,000 times and he has been recognized as the World's Most Influential Scientific Mind, also known as a Highly-Cited Researcher, by Thomson Reuters. He was awarded IEEE Fellow for his contributions to signal processing for wireless communications in 2005. He won 2010 Stephen O. Rice Prize Paper Award, 2013 WTC Wireless Recognition Award, and 2017 Award for Advances in Communication from the IEEE Communications Society and 2013 James Evans Avant Garde Award and 2014 Jack Neubauer Memorial Award from the IEEE Vehicular Technology Society. He also received 2015 Distinguished Faculty Achievement Award from the School of Electrical and Computer Engineering, Georgia Tech. He has been involved in editorial activities for over 20 technical journals for the IEEE, including founding Editor-in-Chief of IEEE 5G Tech Focus. He has organized and chaired many international conferences, including technical program vice-chair of IEEE ICC'03, technical program co-chair of IEEE SPAWC'11, general chair of IEEE GlobalSIP'14, and technical program co-chair of IEEE VTC'16 (Spring).
\end{IEEEbiography}

\begin{IEEEbiography}[{\includegraphics[width=1in,height=1.25in,clip,keepaspectratio]{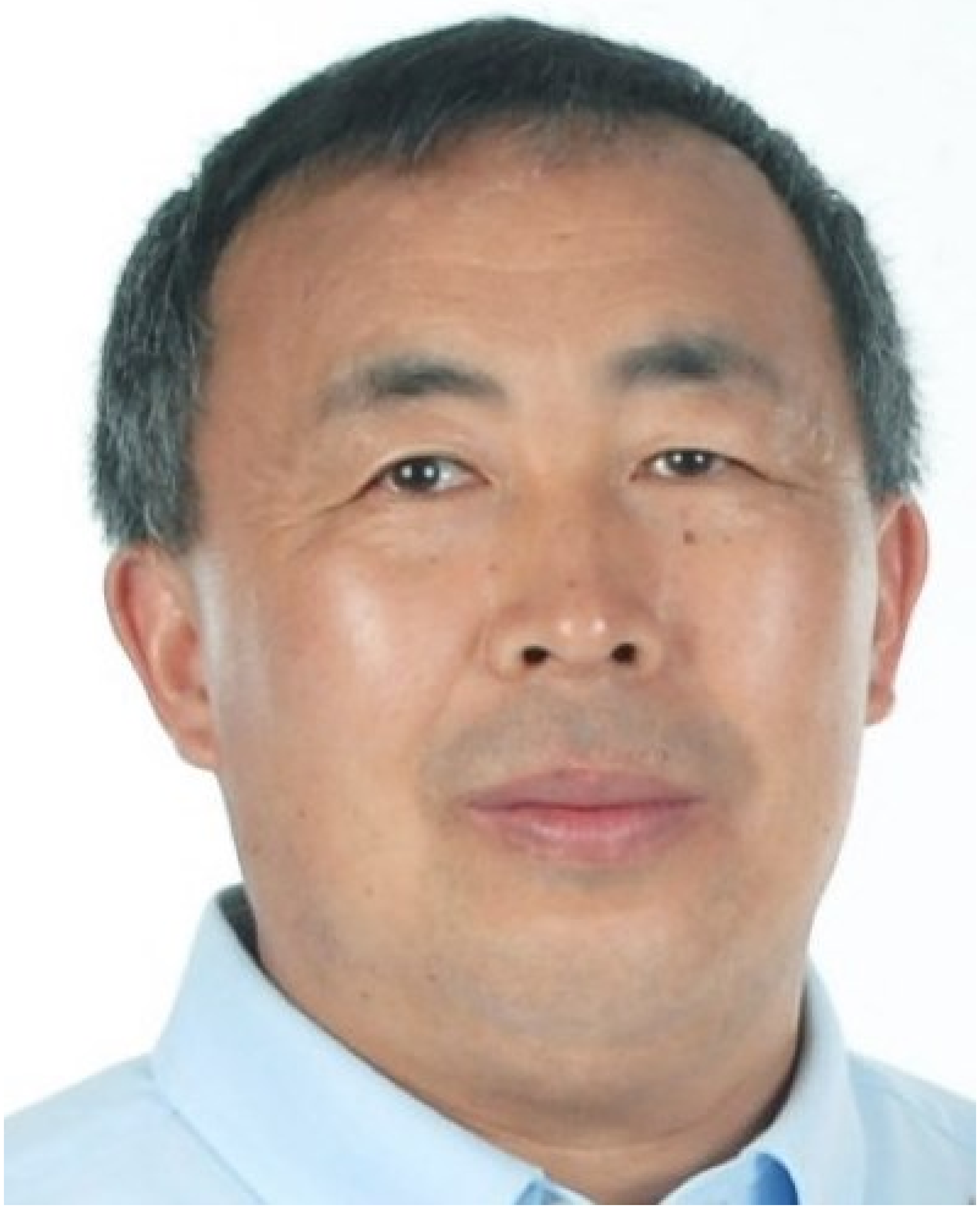}}]{Xuemin (Sherman) Shen}
(M'97-SM'02-F'09) received the B.Sc.(1982) degree from Dalian Maritime University (China) and the M.Sc. (1987) and Ph.D. degrees (1990) from Rutgers University, New Jersey (USA), all in electrical engineering. He is a University Professor and Associate Chair for Graduate Studies, Department of Electrical and Computer Engineering, University of Waterloo, Canada. Dr. Shen's research focuses on resource management in interconnected wireless/wired networks, wireless network security, social networks, smart grid, and vehicular ad hoc and sensor networks. He was an elected member of IEEE ComSoc Board of Governor, and the Chair of Distinguished Lecturers Selection Committee. Dr. Shen served as the Technical Program Committee Chair/Co-Chair for IEEE Globecom'16, Infocom'14, IEEE VTC'10 Fall, and Globecom'07, the Symposia Chair for IEEE ICC'10, the Tutorial Chair for IEEE VTC'11 Spring and IEEE ICC'08, the General Co-Chair for ACM Mobihoc'15, Chinacom'07 and QShine'06, the Chair for IEEE Communications Society Technical Committee on Wireless Communications, and P2P Communications and Networking. He also serves/served as the Editor-in-Chief for IEEE Network, Peer-to-Peer Networking and Application, and IET Communications, and IEEE Internet of Things Journal, a Founding Area Editor for IEEE Transactions on Wireless Communications; an Associate Editor for IEEE Transactions on Vehicular Technology, Computer Networks, and ACM/Wireless Networks, etc.; and the Guest Editor for IEEE JSAC, IEEE Wireless Communications, IEEE Communications Magazine, and ACM Mobile Networks and Applications, etc. Dr. Shen received the Excellent Graduate Supervision Award in 2006. He is a registered Professional Engineer of Ontario, Canada, an IEEE Fellow, an Engineering Institute of Canada Fellow, a Canadian Academy of Engineering Fellow, a Royal Society of Canada Fellow, and a Distinguished Lecturer of IEEE Vehicular Technology Society and Communications Society.

\end{IEEEbiography}

\end{document}